\documentclass[useAMS,usenatbib]{mn2e}
\usepackage{rotating}
\usepackage{txfonts}
\usepackage{graphicx}
\usepackage{multirow}

\usepackage{amssymb}
\usepackage{graphicx}
\def\cmM2{\rm cm$^{-2}$}

\def\1833{PSR~J1833-1034}
 
\def\chandra{{\it Chandra}}
\def\integral{{\it INTEGRAL}}
\def\xmm{{\it XMM-Newton}}

\title[Hard X-ray observations of PSR J1833-1034]{Hard X-ray observations of  PSR J1833-1034 and its associated pulsar wind nebula}
\author[A. De Rosa]{A. De Rosa$^{1}$\thanks{E-mail:
alessandra.derosa@iasf-roma.inaf.it} P. Ubertini$^{1}$, R. Campana$^{2}$, A. Bazzano$^{1}$,  A. J. Dean${^3}$, L. Bassani$^{4}$\\
\\
$^{1}$INAF/IASF-Roma, Via del Fosso del Cavaliere, I-00133 Roma Italy\\
$^{2}$Departement of Physics, University of Rome ''La Sapienza'', Piazzale A. Moro2, I-00185, Roma Italy\\
$^{3}$School of Physics and Astronomy, University of Southampton
Highfield, Southampton, United Kingdom\\
$^{4}$INAF/IASF-Bologna, via Gobetti 101, I-40129 Bologna, Italy
}
\begin{document}

\date{Accepted . Received ; in original form }

\pagerange{\pageref{firstpage}--\pageref{lastpage}} \pubyear{2008}

\maketitle

\label{firstpage}

\begin{abstract}
\1833 and its associated Pulsar Wind Nebula (PWN) has been 
investigated in depth through X-ray observations ranging
from 0.1 to 200 keV.
The low energy X-ray data from \chandra\ reveal a complex morphology that is
characterised by a bright central plerion, no thermal shell and an
extended diffuse halo. The spectral emission from the central plerion
softens with radial distance from the pulsar,
with the spectral index ranging from $\Gamma $ = 1.61  in the central
region to $\Gamma $ =2.36 at the edge of the PWN.
At higher energy \integral\ detected the source in the 17--200 keV range.
The data analysis clearly shows that
the main contribution to the spectral emission in the hard X-ray energy
range is originated from the PWN, while the pulsar  is dominant
above 200 keV.
Recent HESS observations in the high energy gamma-ray domain  show that
\1833 is a bright TeV emitter, with a flux corresponding to $\sim$2 per cent of the Crab in
1--10 TeV range. In addition the spectral shape in the TeV energy region
matches well with that in the hard X-rays observed by \integral.
Based on these findings, we conclude that the emission from the pulsar and its associated PWN can be described in a
scenario where hard X-rays are produced through synchrotron light
of electrons with Lorentz factor $\gamma\sim$10$^{9}$ in a
magnetic field of $\sim$10 micro Gauss. In this hypothesis the TeV emission
is due to Inverse  Compton interaction of the cooled electrons off the
Cosmic Microwave Background photons.  Search for \1833 X-ray pulsed
emission, via RXTE and Swift X-ray observations, resulted in an upper limit that
is  about 50 per cent.

\end{abstract}


\section{Introduction}

G21.5-0.9 is a Crab nebula like supernova remnant (SNR) with a 
plerionic structure, i.e. with a filled center form and  a flat radio spectrum F$_{\nu} \propto \nu^{\alpha}$ with $\alpha$ between 0 and -0.3 (Weiler \& Panagia 1978). It does not display any clear evidence of a thermal shell. Through radio observations, the system was found to include a 61.8 ms pulsar near to the centre of the
nebula (Camilo et al. 2006). The X-ray morphology investigated in depth through several \chandra\, observations (Matheson and Safi-Harb, 2005, Safi-Harb et al. 2001, Camilo et al. 2006) shows a near  spherical structure with a bright compact source surrounded by the
PWN ($\sim $40 arcsec in radius), which is in turn surrounded by non-thermal halo of about 150 arcsec in radius. 
This extended halo shows evidence of a possible SNR structure,
exhibiting filamentary-like features and limb brightening. However, the X-ray
spectral emission from this region is decidedly non thermal in character, as
are the brighter knots within its envelope. 
The origin of this component is still unclear, but its non-thermal nature suggests it is an extension of the synchrotron nebula (Safi-harb et al. 2001), with a contribution of dust scattered X-rays from the plerion (Bocchino et al. 2005).
The detailed X-ray \chandra\, image of the
central zone shows that the compact source, i. e. the site of the pulsar, is surrounded by a
double lobed elliptical emission region of size $\sim $7 arcsec
 by 5 arcsec (Camilo et al. 2006). The offset of the pulsar from the geometrical
centre of the G21.5-0.9 shell is $\lesssim $5 arcsec. The
elliptical geometry is possibly an indicator of a toroidal structure around
the pulsar as observed for the Crab and Vela pulsars or eventually the
putative axisymmetric jet structures observed by F\"{u}rst et al. (1988) by means of  high resolution 22.3-GHz observations of this system. 

The X-ray emission, as observed by \chandra, is well described by a power law model with a spectral
index that is a function of the angular distance from the central region (Safi-Harb et al. 2001).
The point-like X-ray source within a 1 arcsec region at the site of
the pulsar has a spectral index of $\Gamma $ = 1.4. The spectral emission
from the central plerion softens with the radial distance from the pulsar,
and shows a spectral index ranging from a value $\Gamma $ =
1.61, in a region within the central 5 arcsec, to $\Gamma $ =
2.36, at the edge. In  the 0.1--10 keV energy range the outer halo emission
is generally well described by a softer
spectrum with $\Gamma $ $\sim $2.4, while the knots have spectral indices in the
range from $\Gamma \sim $2.14 to $\Gamma $ $\sim $2.5. In 0.5 -- 10 keV energy range, the total unabsorbed flux from the extended component and compact core 
 is 1.1 $\times $10$^{-10}$ erg cm$^{-2}$ s$^{-1}$ and the
corresponding N$_{H}$ value is 2.24 $\times $10$^{22}$ cm$^{-2}$. Following
the detailed discussion presented in Camilo et al. (2006) we adopt a
distance estimate of 4.7 $\pm $ 0.4 kpc.

Careful searches in the\textit{\ Chandra} and XMM data has failed to reveal any
pulsed X-ray emission at the period and ephemeris of the radio pulses. La
Palombara and Mereghetti (2002) set an upper limit on the 0.5 to 10 keV
pulsed fraction between 7.5 per cent and 40 per cent, however, they did not search pulsed events in the range of frequencies that comprises the spin frequency predicted by the radio ephemeris (the pulsed emission was discovered three years later by Camilo and co-workers).
With a period of 61.8 ms and a period derivative of 2.02 $\times $ 10$^{-13}$
s s$^{-1}$, the characteristic spin down age for a braking index of n = 3 is
4.8 ky.  PWN evolution in the light of the spinning neutron star
energy loss rate, suggests  the age should
be much less that 5 ky (Chevalier 2004). There is good
historical evidence from ancient Chinese records to believe that PSR
J1833-1034 is associated with a guest star supernova explosion that took
place in BC 48, making the system just over 2050 years old (Wang et al.,
2006). This younger age implies the pulsar was born with a period 
close to 47 ms.

Finally, very recently HESS telescope has detected Very High Energy (VHE) emission in the TeV energy range from \1833 and its associated PWN (Djannati-Atai et al. 2007).

In this paper we present the  analysis of the \integral-IBIS\, (Ubertini et al. 2003) hard X-ray observation of \1833. The spectral information about this pulsar and its associated PWN above 10 keV was up to now still missing. To search for X-ray pulsation we performed a detailed temporal analysis of the available
archival RXTE and Swift-XRT data.
In Sect. \ref{integral} we investigate the hard X-ray energy spectrum of the pulsar and PWN through  \integral\, observation,  while the TeV HESS emission from \1833 and its association with the hard X-ray  counterpart is discussed in Sect. \ref{hess}. Timing analysis is presented in Sect. \ref{timing} and finally our conclusions are drawn in Sect. \ref{conclusions}.

\begin{figure}
\vspace{-3cm}
\begin{minipage}{10cm} 
\includegraphics[width=1.\linewidth]{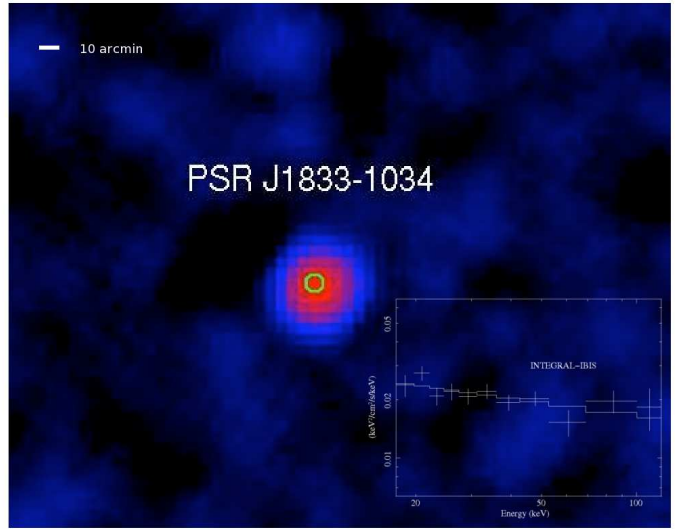}
\end{minipage}
\caption{INTEGRAL image of \1833 in 17--40 keV range. The spatial extension of the excess reveals a point-like source. The circle represents the extension of HESS J1833-1034 (5 arcmin radius). In the insert we show the INTEGRAL spectrum.}
\label{ima}
\end{figure}

\section[]{The hard X-ray spectra of the pulsar and PWN observed by INTEGRAL}
\label{integral} 

\1833 has been observed as part of the third 
\emph{INTEGRAL} IBIS/ISGRI (here after INTEGRAL) survey processing (Bird et al. 2007), which consists 
of all exposures from the beginning of the mission (November
2002) up to April 2006. The total exposure on \1833 is of 
$\sim$ 1.7 Ms.
\integral\, images for each available pointing were generated
in various energy bands using the ISDC offline scientific analysis
software OSA (Goldwurm et al. 2003) version 5.2. The individual images
were then combined to produce a mosaic of the entire sky to enhance
the detection significance using the system described in detail by
Bird et al. (2007). Fig. \ref{ima} shows the 17--40 keV energy band image of the region
surrounding \1833 and  a clear excess is detected at a
significance of $\sim$25$\sigma$. 
The \integral\, image is consistent with a point-like source that is coincident
with the site of PSR J1833-1034. The data are well fitted by a simple power-law with photon index $\Gamma $ = 2.2$^{+0.1}_{-0.1}$ (the spectrum is shown in the insert in Fig. \ref{ima}). The 20--100 keV flux of 5.2$\times$10$^{-11}$ erg cm$^{-2}$ s$^{-1}$, as measured by \integral, comprises 0.4 per cent of the pulsar spin down luminosity of $\dot{E}$ = 3.38 $\times $10$^{37}$ erg s$^{-1} $, assuming a distance to the source of 4.7 kpc. The most likely
site of this emission should be from the central bright plerion. \chandra\, observations in 0.1--10 keV energy range have clearly detected the softening of the X-ray
emission with radial distance from the pulsar, that is presumably the source of
energetic electrons. This evidence has all the hallmarks of synchrotron cooling, so that
we would expect the hard X-rays as observed by \integral\, to originate nearby the central source.  

\begin{figure}
\centering
\begin{minipage}{9cm} 
\includegraphics[width=1.0\linewidth]{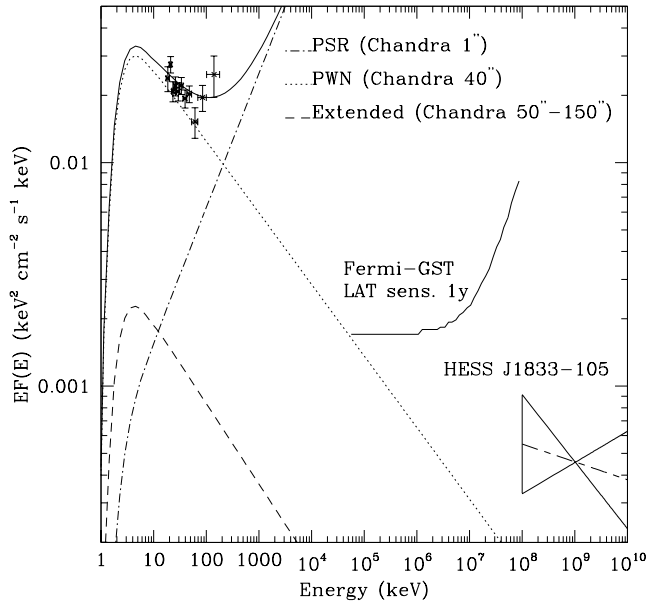}
\end{minipage}
\caption{Spectral Energy Distribution from soft X-ray to TeV energy range. Each spectral component identified through \chandra\, observation has been plotted separately. 
\integral\, data are plotted on the top of the model. We also plot the Fermi-GST (formerly GLAST) LAT 1-year sensitivity curve taking into account the diffuse Galactic and the residual instrumental background.}
\label{spec}
\end{figure}

To evaluate the contribution to the \integral\, spectrum of the different components of the source (pulsar, PWN and extended emission), we plot in Fig. \ref{spec} for each component  the spectral model as observed by \chandra\, (Safi-Harb et al. 2001).  The low surface brightness extended emission (originated between 50 arcsec and 150 arcsec radius around the pulsar), is reproduced with a soft power-law ($\Gamma$=2.36, dashed line). 
The emission of the pulsar (dot-dashed curve), extracted from the central innermost region in a circle with 1 arcsec radius, is reproduced with a hard power-law ($\Gamma$=1.4). The emission of the PWN (dotted line) is extracted from  a circle of 40 arcsec radius and is best fitted by a softer power-law with respect to the pulsar ($\Gamma$=2.3). All components are absorbed by a cold gas with  column density N$_{\rm H}$=2.2$\times$10$^{22}$cm$^{-2}$  as derived through \chandra\, (Safi-harb et al. 2001) and \xmm\, (Warwick et al. 2001) observations.
In Fig. \ref{spec} the \integral\, data are also shown.
From this figure it appears the pulsar contributes very little  in the 20--200 keV energy range, and the extended emission is too weak with respect to the other components to affect  the high energy emission as detected by \integral. Therefore the PWN is likely to  contribute most to the hard X-ray flux. 
Indeed, the contribution of the pulsar at the 20--100 keV flux, as derived through \integral\, data, is lower than 20 per cent and it become even lower in the soft X-ray energy range, i.e. $\sim$ 3 per cent (see also Kargaltsev \& Pavlov 2008). This is clearly visible in the SED plotted in Fig. \ref{spec}. In 20--100 keV energy range a fraction of the total emission that is larger of 80 per cent is originated in the PWN.
Using the pulsar spin down luminosity of $\dot{E}=3.38\times 10^{37}$ erg s$^{-1}$, the total energy conversion efficiency (in 20--100 keV energy range) is $\eta_{tot}=L_{tot}/\dot{E} \sim$0.4 per cent, with the PWN contribution that is $L_{PWN}/\dot{E}\sim$ 0.3 per cent and the pulsar $L_{pulsar}/\dot{E}\sim$0.1 per cent.
The value of the total  energy conversion efficiency $\eta_{tot}$ is well in agreement with the average value measured in the others pulsar/PWM systems observed by \integral, that is around 1 per cent (PSR~J1846-0258, McBride et al., 2008; PSR~J1617-5055, Landi et al., 2007; Vela, Hoffmann et al., 2007; PSR~J1513-5906, Forot et al., 2006; PSR~J1811-1925, Dean et al., 2008). PSR B0540-69 (Campana et al. 2008) is characterized by the highest value $\eta_{tot}\sim5.8$ per cent, while Vela by the lowest  $\eta_{tot}\sim 0.02$ per cent.
Finally we note that the ratio between the energy conversion efficiency of the PWN and pulsar is $\eta_{PWN}/\eta_{PSR}\sim$ 4.3, in very good agreement with the mean value ($\sim$4) for a sample of X-ray pulsar/PWN systems observed in the soft X-ray energy range with \chandra\, (Kargaltsev \& Pavlov 2008).

\begin{table*}
\begin{flushleft}
\caption{Hard X/gamma-ray properties of \1833 and its PWN detected by \integral\, and HESS}
\begin{tabular}{ccccccc}
\noalign{\hrule}
F$_{20-100 \,\rm keV}$ & $^\star$L$_{\rm X}$(\%$\dot{E}$) & $\Gamma_{\rm X}$ & F$_{\rm PWN}$/F$_{\rm PSR} $ & $^{\dagger}$L$_{\gamma}$(\%$\dot{E}$) & $\Gamma_{\gamma}$ &  L$_{\rm X}$/L$_{\gamma}$ \\
 (10$^{-11}$erg cm$^{-2}$s$^{-1}$) &  (10$^{34}$erg s$^{-1}$) & & at 10 keV & (10$^{34}$erg s$^{-1}$) & & \\
\hline
5.2 & 13(0.4) & 2.2$\pm$0.1& 10  &  0.4(0.013)& 2.1$\pm$0.2 & 30 \\

\hline
\end{tabular}
\label{property}

\small{$^\star$ In 20-100 keV. $^{\dagger}$ In 1-10 TeV.}
\end{flushleft}
\end{table*}
\section[]{association with HESS J1833-105}
\label{hess} 

Observations in the TeV energy range have demonstrated that  \1833 and its associated PWN are powerful emitters in the VHE gamma-rays (Djannati-Atai et al. 2007).
The position of HESS J1833-105 (18$^{h}$33$^{m}$32.5$^{s}\pm 0.9^{s}$, -10d33'19$^{\shortparallel }\pm 55^{\shortparallel }$) is in very good agreement with the  location  of the source detected by \integral. The properties of the hard X/gamma-ray emission from the pulsar/PWN system observed by \integral\, and HESS are reported in Table \ref{property}, while in Fig. \ref{ima} the HESS position is shown as a circle on the \integral\, image. The photon index of the TeV spectrum is 2.1$\pm$0.2 with a flux corresponding to $\sim$2 per cent of the Crab (see Table \ref{property}).
In Fig. \ref{spec} we report the best-fit power-law reproducing the spectrum of HESS J1833-105, with the three lines representing the upper, lower and best fit value of the slope. The shape of this emission, extrapolated in the X-ray energy range, matches well with the PWN spectrum observed by \integral, with the ratio between the X-ray and gamma-ray luminosity of L$_{X}$/L$_{\gamma}\sim$ 30.
We assume that the soft-gamma ray emission is produced through synchrotron process of  electrons in a magnetic field of 10 micro Gauss. The value of the magnetic field is close to that implied from gamma-ray observations assuming  (a) the gamma-ray flux is produced by Inverse Compton (IC) scattering on Cosmic Microwave Background (CMB) photons and  (b)  a ratio  L$_{X}$/L$_{\gamma}\sim$ 30 (Djannati-Atai et al. 2007). The equipartition value suggests the presence of a magnetic field much higher, of about B=0.3 mG (Camilo et al. 2006). However, HESS observations seems to suggests that the magnetic field in PWN can be much lower than the equipartition value (Chevalier 2004).
To produce the observed  20--100 keV emission through synchrotron light in 10 micro Gauss magnetic field, we need electrons with Lorentz factor of $\sim$10$^{9}$. This value remains still very high even if we assume a higher magnetic field, e.g. $\gamma$ $\sim$10$^{8}$ for B=50 micro Gauss. In this scenario TeV emission can be produced through IC scattering of the same electrons population with  CMB light.
For electrons with  Lorentz factor of $\sim$10$^{9}$, the maximum energy gained in a single scattering is $\nu_{max}=\gamma^{2}\nu_{CMB}\sim 10^{14}$ eV. The energy of the electrons producing TeV emission can be lower with respect to that required to produce the 20--100 keV spectrum. This matches well with the hypothesis that the TeV emission is produced by the electrons cooled through synchrotron radiation in a 10 micro Gauss field. 

Future Fermi-GST observations will allow a detailed study of the SED in the MeV--GeV energy range, where the pulsar contribution should be dominant. The 1 year LAT sensitivity curve is also plotted in Fig. \ref{spec}.

In the sample of pulsar/PWN systems detected by \integral, \1833 together with Crab and PSR J1846-0258, is characterised by a high L$_{20-100 keV}$/L$_{1-10 TeV}$ ratio and  by positional coincidence with HESS.
This high ratio seems to be an artefact of young pulsar/PWN systems for which the TeV source is coincident with the pulsar, and maybe due to the fact that the electrons (synchrotron cooled) reservoir is not completely filled because of the lack of accumulation time.

\section[]{Timing analysis}
\label{timing}
In the case of \integral\, data, the combination of the cumulative long observation period and the
relatively low statistical significance of the detection make impossible
to isolate and measure a pulsed hard X-ray component in the emitted
flux.
We then performed a period search using RXTE/PCA data in order to establish an upper limit on the pulsed emission in the range 2--10 keV.
PSR J1833-1034 has been observed once by RXTE, during proposal ID 20259 performed in November 8, 1997 with G21.5-0.9 as a target. We have selected PCA Good-Xenon observations, i.e. the data mode with the full timing resolution. The data were filtered using the standard criteria (elevation angle, south atlantic anomaly passages, etc); the baricentrisation was performed using the task \texttt{fxbary} and the pulsar location from \chandra\, (Camilo et al. 2006). The total exposure time on the source was 73 ks. We performed a period search in a reasonably large frequency range around the
extrapolation of the radio ephemeris to the RXTE observation time, since the radio ephemeris were not available at the time of RXTE observations.
We used the $Z^2_m$ test (Buccheri et al. 1993) with $m = 2$ and $m = 10$ harmonics. 
We ignored the $Z^2_m$ test with $m=1$ because this is very powerful for sinusoidal peaks, while we have no idea on lightcurve expected in our case (Protheroe 1987).
We found no pulsation at the 99 per cent confidence level.
Upper limits on the pulsed emission were derived with the hypothesis that the pulse shape is
either sinusoidal or very narrow (in the limit of a Dirac delta shape), following the
prescriptions of Protheroe (1987) and De Jager et al. (1989). For the 2--10 keV band we obtained a flux value between  2.5 and 1.2$\times$10$^{-5}$ photons cm$^{-2}$s$^{-1}$ for a sinusoid and a Dirac delta pulse,
respectively. This pulsed flux upper limit corresponds to 1.7$\times$ 10$^{-13}$ erg cm$^{-2}$s$^{-1}$ and 8.8$\times$10$^{-14}$
erg cm$^{-2}$s$^{-1}$ respectively, assuming that the pulsar photon index is  $\Gamma = 1.5$ as found by \chandra. These upper limits are of the same order of magnitude, if not slight more stringent, than those set by \chandra\, (Camilo et al. 2006). The pulsed emission is therefore
lower than 50 per cent of the total emission from the pulsar.

We also analysed observations of PSR J1833-1034 made with the XRT telescope onboard the Swift
satellite (Gehrels et al. 2004). We selected all the observations having this source in the field of view and performed
in the Windowed Timing (WT) mode, that ensures the full timing capability of this instrument, with
a time resolution of about 1.7 ms. The dataset contains 9 pointings, performed between October 2
and 21, 2007.  The total net exposure time after the standard pipeline filtering is about 52 ks.
We performed a period search using an approach similar to the one used for RXTE observation, and still no significant pulsation was found at the 99 per cent confidence level. The upper limit on the pulsed flux was of $\sim$5$\times$10$^{-13}$ erg cm$^{-2}$s$^{-1}$.

\section{Conclusions}
\label{conclusions} 

The combined spectral study of the Crab-like pulsar \1833 and its associated PWN over the \chandra\, and \integral\, energy range, strongly suggests that the 20--100 keV flux  is mainly due to the PWN component (more than 80 per cent of the total flux), and the PSR contribution can only be dominant above 200 keV. 
\1833 is also a bright TeV emitter as detected by HESS. The \integral\, emission represents the hard X-ray counterpart to the VHE HESS source and the TeV emission matches well with the 17--200 keV spectral shape.
But for PSR~J1617-5055 and PSR~J1513-5906, most of the pulsar/PWN systems observed by \integral\, have the hard X-ray emission dominated by the PWN. However, a detailed comparison between these sources  shows that the relative contribution from the pulsar and the PWN differs very much from source to source.
The observed properties of \1833 suggest the hard X-ray  emission is produced through synchrotron light of electrons with $\gamma\sim$10$^{9}$ in a magnetic field of $\sim$10 micro Gauss while the TeV emission is due to IC of the cooled electrons on the CMB photons. 

We note that in the sample of pulsar/PWN systems observed by \integral\, \1833 together with
Crab and IGR J1846-0258 are the only systems showing  similar properties, i.e. high L$_{\rm 20-100 keV}$/L$_{\rm 1-10 TeV}$ ratio ($\gg$1 ) and with the position of the HESS coincident with that of the pulsar/PWN. This evidence suggests that the cooled electrons responsible for the TeV emission remain confined in the same region of the synchrotron emission. This could be an artefact of the young pulsar/PWN system.

We did not detected any evidence of pulsations in RXTE and Swift-XRT data (with upper limit for the pulsed emission of 50 per cent of the total). The lack of X-ray pulsed emission is not fully unexpected in view of the moderate magnetic field of the pulsar, B=3.58$\times$10$^{12}$ Gauss. However, we stress also that Crab and PSR B0540-69, clearly exhibiting X-ray pulsations, are characterized by a magnetic fields of 3.78$\times$10$^{12}$  and 4.96$\times$10$^{12}$ Gauss respectively, i.e. values similar to that observed in \1833.
The lack of X-ray pulsed emission from \1833 could also be explained with the radio and X-ray beams  pointing to different directions or with a different aperture, and the line of sight sweeps only one of them. A similar hypothesis has been proposed to explain the behaviour of Geminga and the radio-quiet pulsars (Harding et al. 2007). \\


This program is funded by Italian Space Agency  grant  via contracts I/008/07/0 and I/088/06/0.

\bsp

\label{lastpage}

\end{document}